\documentclass[12pt]{article}
\usepackage{geometry}
\usepackage{subfigure}
\usepackage{physics}
\usepackage{multirow}
\usepackage{booktabs}
\numberwithin{equation}{section}
\makeatletter
\newcommand\ackname{Acknowledgements}
\if@titlepage
\newenvironment{acknowledgements}{
	\titlepage
	\null\vfil
	\@beginparpenalty\@lowpenalty
	\begin{center}
		\bfseries \ackname
		\@endparpenalty\@M1
\end{center}}
{\par\vfil\null\endtitlepage}
\else
\newenvironment{acknowledgements}{
	\if@twocolumn
	\section*{\abstractname}
	\else
	\small
	\begin{center}
		{\bfseries \ackname\vspace{-.5em}\vspace{\z@}}
	\end{center}
	\quotation
	\fi}
{\if@twocolumn\else\endquotation\fi}
\usepackage{mathrsfs}
\usepackage{eucal}
\usepackage{amssymb,amsmath}
\usepackage{pgf}
\usepackage{tikz}
\usepackage{cite}
\usepackage{graphicx}
\usepackage{color}
\usepackage{amssymb}
\usepgflibrary{arrows}
\usetikzlibrary{calc,angles,positioning,intersections,quotes,backgrounds,decorations.pathreplacing}
\usetikzlibrary{decorations.markings}
\usepackage{amsmath}
\usepackage{amsthm}
\usepackage{accents}

\theoremstyle{remark}
\theoremstyle{definition}

\usepackage[makeroom]{cancel}

\newcommand{\nn}{\nonumber}
\usepackage{comment}
\begin{document}
	\title{\textbf{The action principle for equilibrium thermodynamics}}
	\author{Sikarin Yoo-Kong \\	
		\small  \emph{The Institute for Fundamental Study (IF),} \\
		\small\emph{Naresuan University (NU), Phitsanulok-Nakhon Sawan,}\\
		\small  \emph{99 Moo 9, Tha Pho, Mueang Phitsanulok, 65000 Phitsanulok, Thailand.}	\\
        \small\emph{e-mail:\;sikariny@nu.ac.th}
	}
	\date{}
	\maketitle
	\abstract
	The action principle is introduced to describe the thermodynamic processes of the state functions from the initial equilibrium state to the final equilibrium state. %The Lagrangian 1-form formalism is introduced to exhibit the exactness, namely the Maxwell’s relations, of the thermodynamic functions.
 To capture the path-independent property of the state functions through the thermodynamic processes, one requires an integrability condition called the Lagrangian 1-form closure relation as a direct result of the least action principle with respect to the independent variables.
	
	\section{Introduction}\label{S1}
	It is well known that, in the standard equilibrium thermodynamics, the evolution of the thermodynamic process is expressed on the thermodynamic phase space or phase diagram, e.g. $PVT$ or $TPS$. On this particular space, the equation of state (EOS) is defined as a point and the evolution of the state functions, e.g. internal energy and entropy, from one equilibrium point to another equilibrium point is path-independently given. On the contrary, in the context of classical mechanics, the equation of motion (EOM) is presented to describe the dynamics of the system on the tangent bundle or contangent bundle. The action functional plays a major role and the least action principle demands the action to be critical resulting in a classical path. Now it comes to an interesting question: Is there a way to formulate the action principle in the context of thermodynamics? Up to our humble knowledge, the answer is not that yet. However, there were many attempts, dated back to the early period of formulating thermodynamics as well as statistical mechanics, to capture thermodynamic process in terms of Lagrangian mechanics by Hermann Helmholtz \cite{Helm1,Helm2}, Henri Poicare, Pierre Duhem \cite{D1,D2,D3,D4,D5}, Joshua Gibbs \cite{Gibbs1,Gibbs2,Gibbs3}, Joseph John Thomson \cite{Thomson} and even James Clerk Maxwell \cite{M1,M2}. Later, of course, many interesting pieces of work existed in the literature and we cannot mention all of them. However, it is worth to mention some interesting works. The first one is from Szamosi on ``variational principle in thermodynamics" \cite{GS} to express the first law of thermodynamics from the Euler-Lagrange equation. The second one is from Buchdahl on ``A variational principle in classical thermodynamics" \cite{B} to formulate the second law of thermodynamics from the variational principle. The third one is ``Contact geometry and thermodynamics" by Bravetti \cite{Bravetti} to capture thermodynamic processes in the geometrical language and the last one is on ``A Hamiltonian approach to Thermodynamics" by Baldiotti and his colleagues to make an analogy between thermodynamic systems with the Hamiltonian one \cite{Baldiotti}.
 \\
 \\
 In this work, we shall construct the action principle to explain the thermodynamic processes that take state functions from an initial equilibrium point to a final equilibrium point on the phase diagram. A major point is to capture the path-independent feature of the state functions by employing a new paradigm on Lagrangian mechanics from the context of integrable systems known as Lagrangian 1-form structure and its closure relation \cite{Frank}. The rest of the paper is organised as follows. In section \ref{S2}, the thermodynamics in the context of differential forms will be briefly provided and Maxwell relation will consequently derived. In section \ref{LI}, the action principle will be constructed in terms of Lagrangian 1-forms. The exactness of the state functions will be disguisedly expressed through the Lagrangian closure relation as a direct result on the variation of independent variables. In section \ref{SS}, the summary together with some remarks will be mentioned.
 \section{Differential forms and Maxwell relations}\label{S2}
\begin{figure}[h]
\centering
\includegraphics[width=10cm]{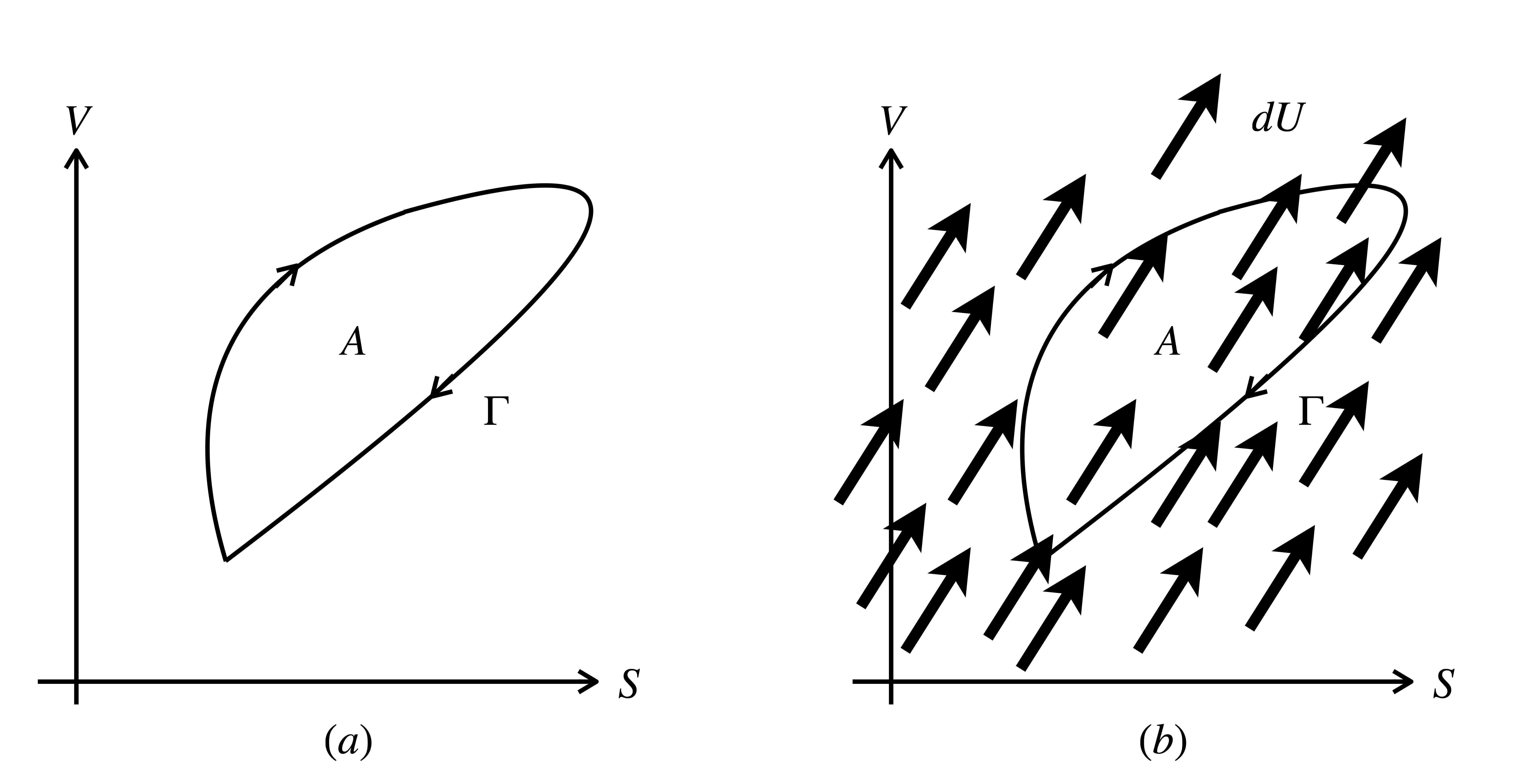}
\caption{(a): A reversible loop process on the $S-V$ plane. (b): The change of the vector field $dU$ must be zero. }\label{F1}
\end{figure}

In this section, we shall present the thermodynamic potentials in the differential form and shall also derive all Maxwell relations. We first note here that this scheme on differential forms for thermodynamics is not new but was investigated long time ago \cite{MRUGALA}. However, to make this present paper self-contained, we shall give a brief review.
\\
\\
To make thing simple, we shall work with an internal energy $U$ expressed as
\begin{equation}\label{U1}
    dU=TdS-PdV\;,
\end{equation}
where $T$ is a temperature, $S$ is an entropy, $P$ is a pressure and $V$ is a volume. We note that $\{S,V\}$ will be treated as a set of independent variables, namely $U=U[S,V]$. With the present form of \eqref{U1}, one can easily see that $dU$ is nothing but the 1-form. What we are interested is the integration such that 
\begin{equation}
    \oint_\Gamma dU=\oint_\Gamma \left(TdS-PdV\right)\;,
\end{equation}
where $\Gamma$ is a closed loop defined on the S-V surface in figure \ref{F1}(a). Since, the internal energy $U$ is an extensive thermodynamic quantity, therefore
\begin{eqnarray}\label{U2}
    0&=&\oint_\Gamma dU=\oint_\Gamma \left(TdS-PdV\right)\nn\\
    \oint_\Gamma TdS&=&\oint_\Gamma PdV\;.
\end{eqnarray}
This equation is telling us that the heat added into the system will be converted to be work done by the system along the loop $\Gamma$. Next, applying the Green's theorem, we obtain
\begin{eqnarray}\label{U3}
    \iint_A dT\wedge dS&=&\iint_A dP\wedge dV\nn\\
    \iint_A \omega&=&\iint_A \left(dT\wedge dS-dP\wedge dV\right)=0\;,
\end{eqnarray}
where $A$ is an area bounded by $\Gamma$ and $\omega$ is a 2-form. Interestingly, in the geometrical view, the object $dU$ can be treated as a vector field and \eqref{U2} indicates that the change of the $dU$ along the loop $\Gamma$ is zero, see figure \ref{F1}(b) or, equivalently, $dU$ depends on the end-points, but not the path between these two end-points. This feature known as the path-independent feature.
\\
\\
Next, we shall derive all the Maxwell relations. 
%Recalling \eqref{U3}, we have 
%\begin{equation}
%    dT\wedge dS=dP\wedge dV\;.,
%\end{equation}
%with the antisymmetric property $dT\wedge dS=-dS\wedge dT$. 
Here suppose that $\{ T,S\}$ are function in $\mathbb{R}^d$, where $d \geq 1$, with a set of basis $\{ dx^1,dx^2,...,dx^d \}$. Therefore, we write \eqref{U3} in the form
\begin{equation}
    \frac{dT}{dx^i}\frac{dS}{dx^j}dx^i\wedge dx^j=\frac{dP}{dx^i}\frac{dV}{dx^j}dx^i\wedge dx^j=-\frac{dP}{dx^j}\frac{dV}{dx^i}dx^i\wedge dx^j\;.
\end{equation}
In the case $d=2$ with $\{x^1=X,x^2=Y\}$ which is a set of any variables, one obtains the following relations.
%and we write
%\begin{equation}
%    \frac{dTdS}{dXdY}=\frac{dPdV}{dXdY}\;,
%\end{equation}
%where $\{X,Y\}$ is a set of any variables.
\\
\\
Case 1. If we choose $\{X=V,Y=S\}$, we obtain
\begin{eqnarray}\label{MW1}
    \frac{dT\cancel{dS}}{dV\cancel{dS}}dV\wedge dS&=&\frac{dPdV}{dVdS}dV\wedge dS=-\frac{dP\cancel{dV}}{dS\cancel{dV}}dV\wedge dS\nn\\
    \frac{dT}{dV}&=&-\frac{dP}{dS}\;.
\end{eqnarray}
Case 2. If we choose $\{X=T,Y=V\}$, we obtain
\begin{eqnarray}\label{MW2}
    \frac{\cancel{dT}dS}{\cancel{dT}dV}dT\wedge dV&=&\frac{dP\cancel{dV}}{dT\cancel{dV}}dT\wedge dV\nn\\
    \frac{dS}{dV}&=&\frac{dP}{dT}\;.
\end{eqnarray}
Case 3. If we choose $\{X=P,Y=T\}$, we obtain
\begin{eqnarray}\label{MW3}
    \frac{dTdS}{dPdT}dP\wedge dT&=&-\frac{\cancel{dT}dS}{\cancel{dT}dP}dP\wedge dT=\frac{\cancel{dP}dV}{\cancel{dP}dT}dP\wedge dT\nn\\
    -\frac{dS}{dP}&=&\frac{dV}{dT}\;.
\end{eqnarray}
Case 4. If we choose $\{X=P,Y=S\}$, we obtain
\begin{eqnarray}\label{MW4}
    \frac{dT\cancel{dS}}{dP\cancel{dS}}dP\wedge dS&=&=\frac{\cancel{dP}dV}{\cancel{dP}dS}dP\wedge dS\nn\\
    \frac{dT}{dP}&=&\frac{dV}{dS}\;.
\end{eqnarray}
With this scheme, we successfully derive all Maxwell relations.
\section{Lagrangian 1-form structure and path-independent feature}\label{LI}
\begin{figure}[h]
\centering
\includegraphics[width=10cm]{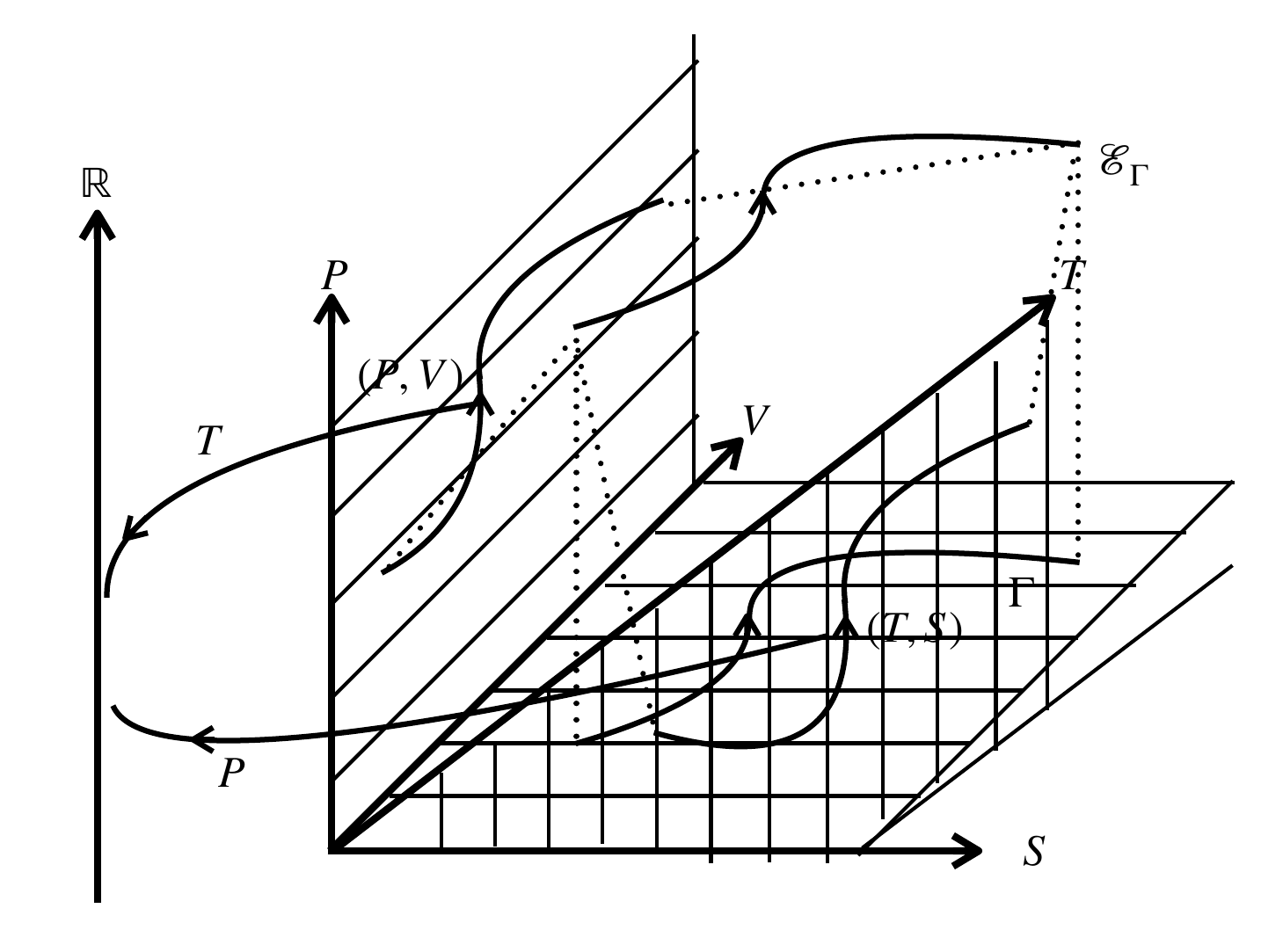}
\caption{The projection of the curve $\mathscr{E}_\Gamma$ on the some particular planes. }\label{phaseS}
\end{figure}

\begin{figure}[h]
\centering
\includegraphics[width=6cm]{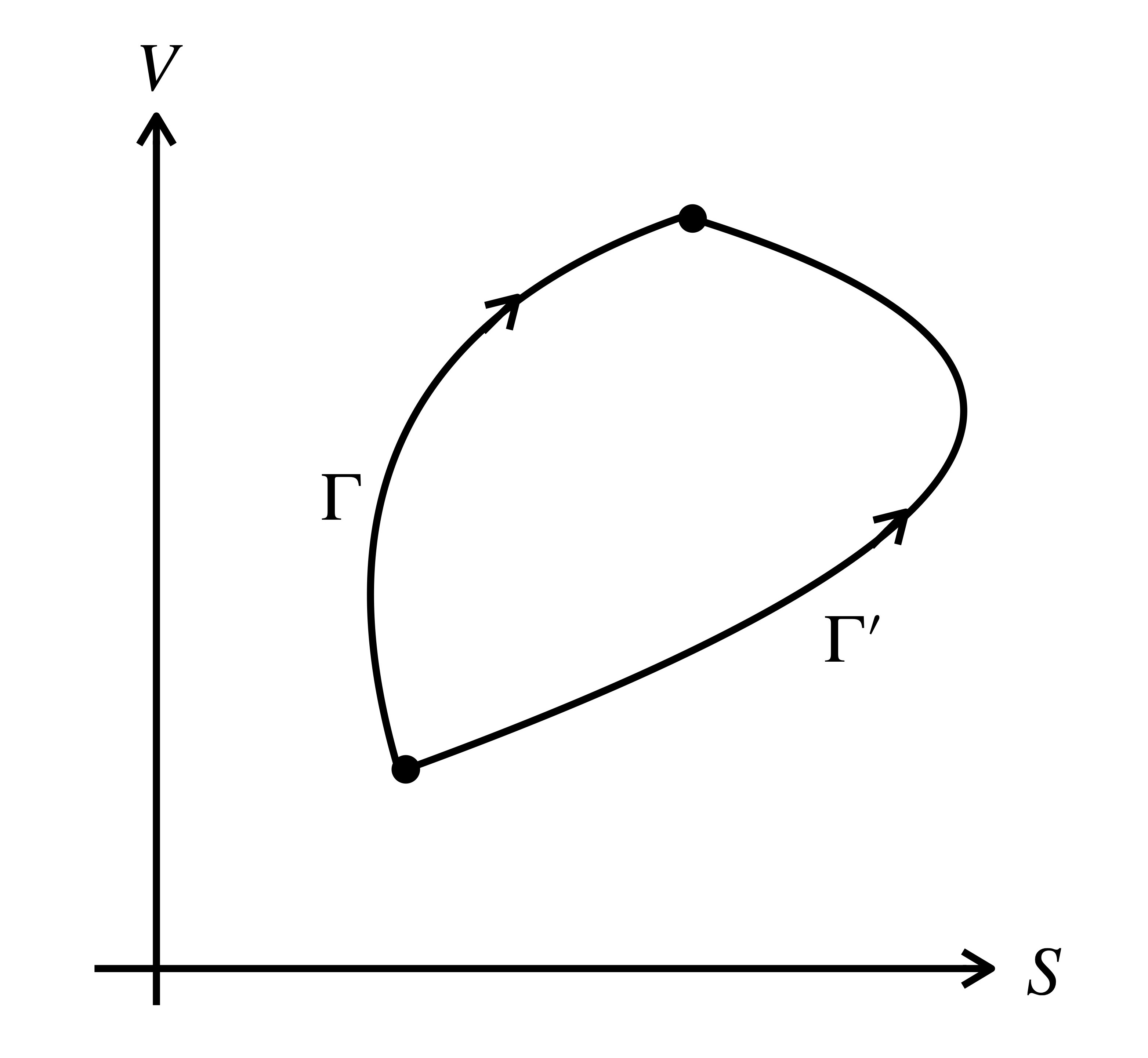}
\caption{The variation of the path on the space of independent variables: $\Gamma\rightarrow\Gamma'$. }\label{F2}
\end{figure}

In this section, we shall formulate the Lagrangian description to capture the exactness or path-independent feature of the state functions. Again, for simplicity and continuity, we still consider the internal energy.  To construct the action, we shall take a close look at the expression of $dU$ and it is not difficult to see that one can write
\begin{equation}\label{du1}
    dU=\left(T\frac{dS}{dV}-P(T,S)\right)dV\;,
\end{equation}
or
\begin{equation}\label{du2}
    dU=\left(T(P,V)-P\frac{dV}{dS}\right)dS\;.
\end{equation}
It is well understood that the change of the internal energy $dU$, both \eqref{du1} and \eqref{du2}, would amount to a specific choice of thermodynamic process.
Here we note that, in \eqref{du1}, the variable $V$ plays a role of the independent variable, and the variable $S$ will be treated as coordinate variable (the dependent variable). Alternatively, if we choose the express $dU$ in \eqref{du2}, the variable $S$ plays a role of independent variable and $V$ will be treated as the coordinate variable (the dependent variable).
Then it seems to suggest that one can take the internal energy as an action functional
\begin{equation}
    U[S,V]=\int_\Gamma L\;,\label{US}
\end{equation}
where $\Gamma$ is a reversible path defined on $S-V$ surface and $L$ is a Lagrangian 1-form given by
\begin{equation}
    L=\mathcal{L}_VdV+\mathcal{L}_SdS\;.
\end{equation}
Here $\mathcal{L}_V$ and $\mathcal{L}_S$ are given by
\begin{eqnarray}
    \mathcal{L}_V\left(\frac{dS}{dV},S;V\right)&=&T\frac{dS}{dV}-P(T,S)\;,\label{LV}\\
    \mathcal{L}_S\left(\frac{dV}{dS},V;S\right)&=&T(P,V)-P\frac{dV}{dS}\;\label{LS}.
\end{eqnarray}
We note here that, actually, the action functional \eqref{US} is a change of the internal energy along the path $\Gamma$. Moreover, if one recalls the standard Lagrangian for a point particle $L(\frac{dq}{dt},q,t)=p\frac{dq}{dt}-H(p,q)$, we find that, for the Lagrangian \eqref{LV}, $T$ plays a role of the momentum variable, $S$ is nothing but a coordinate variable, $V$ will be treated as a time variable and $P(T,S)$ can be viewed as the Hamiltonian. Moreover, the function $P(T,S)$ associated with the Lagrangian in \eqref{LV} can defined on the projection plane $T-S$, shown in figure \ref{phaseS}. The trajectory on this plane can be treated as a projection of the curve $\mathscr{E}_\Gamma$. Similarly, for the Lagrangian \eqref{LS}, $P$ is treated as a momentum variable, $V$ is a coordinate variable, $S$ is a time variable and $T(P,V)$ can be viewed as the Hamiltonian, see also figure \ref{phaseS}.
\\
To capture the exactness feature, see appendix, we consider the variation with respect to independent variables as follows: $V\rightarrow V+\delta V $ and $S\rightarrow S+\delta S $ resulting in a new path $\Gamma'$ with the same end-points, see figure \ref{F2}. 
\\
A new action is given by
\begin{eqnarray}\label{U4}
    U[S+\delta S,V+\delta V]&=&\int_{\Gamma'}\mathcal{L}_V(V+\delta V,S+\delta S)d(V+\delta V)+\mathcal{L}_S(V+\delta V,S+ \delta S)d(S+\delta S)\;,\nn\\
    &=&\int_{\Gamma}\mathcal{L}_VdV+\mathcal{L}_SdS+\int_{\delta\Gamma}\left(\frac{\partial\mathcal{L}_V}{\partial V}dV+\frac{\partial\mathcal{L}_V}{\partial S}dS+\frac{\partial\mathcal{L}_S}{\partial V}dV+\frac{\partial\mathcal{L}_S}{\partial S}dS\right)\nn\\
    &&+\int_{\delta\Gamma}\mathcal{L}_Vd\delta V+\mathcal{L}_Sd\delta S\;.
\end{eqnarray}
Integrating by parts the last two terms, \eqref{U4} becomes
\begin{eqnarray}
    U[S+\delta S,V+\delta V]&=& U[S,V]+\left(\mathcal{L}_V\delta V+\mathcal{L}_S\delta S\right)\Big\vert_{\text{boundary}}-\int_{\delta\Gamma}d\mathcal{L}_V\delta V+d\mathcal{L}_S\delta S\nn\\  &&+\int_{\delta\Gamma}\left(\frac{\partial\mathcal{L}_V}{\partial V}dV+\frac{\partial\mathcal{L}_V}{\partial S}dS+\frac{\partial\mathcal{L}_S}{\partial V}dV+\frac{\partial\mathcal{L}_S}{\partial S}dS\right)\;.
\end{eqnarray}
The boundary terms vanish since end-points are being fixed. Next we employ the fact that
\begin{eqnarray}
    d\mathcal{L}_V&=&\frac{\partial \mathcal{L}_V}{\partial V}dV+\frac{\partial \mathcal{L}_V}{\partial S}dS\;,\\
     d\mathcal{L}_S&=&\frac{\partial \mathcal{L}_S}{\partial V}dV+\frac{\partial \mathcal{L}_S}{\partial S}dS\;.
\end{eqnarray}
Then what we have now is that
\begin{eqnarray}
    \delta U\equiv U[S+\delta S,V+\delta V]-U[S,V]=\int_{\delta\Gamma}\delta V\left(\frac{\partial\mathcal{L}_V}{\partial S}-\frac{\partial\mathcal{L}_S}{\partial V}\right)dS+\delta S\left(\frac{\partial\mathcal{L}_S}{\partial V}-\frac{\partial\mathcal{L}_V}{\partial S}\right)dV\;.
\end{eqnarray}
The least action principle demands that $\delta U=0$ resulting in
\begin{equation}\label{CL1}
    \frac{\partial\mathcal{L}_V}{\partial S}=\frac{\partial\mathcal{L}_S}{\partial V}\;,
\end{equation}
which is called the Lagrangian 1-form closure relation. The relation \eqref{CL1} ensures that under the local deformation of the curve on the space of independent variables, e.g. $\{S,V\}$, the change of internal energy remains the same value. Therefore, this is nothing but the path-independent feature of the state functions in thermodynamics under the evolution. Equivalently, the choice of the thermodynamic processes for \eqref{du1} and \eqref{du2} does not affect \eqref{CL1} reflecting the integrability of $dU$.
\\
\\
\textbf{Theorem}: The closure relation 
\begin{equation}
    \frac{\partial\mathcal{L}_V}{\partial S}=\frac{\partial\mathcal{L}_S}{\partial V}\;,\nn
\end{equation}
holds on the Maxwell relation \eqref{MW1}.
\\
\\
\begin{proof} 
\begin{eqnarray}
    \frac{\partial}{\partial S}\left(T\frac{dS}{dV}-P\right)&=&\frac{\partial}{\partial V}\left(T-P\frac{dV}{dS}\right)\nn\\
    \frac{dT}{\cancel{dS}}\frac{\cancel{dS}}{dV}+T\frac{d^2S}{dSdV}-\cancel{\frac{dP}{dS}}&=&\cancel{\frac{dT}{dV}}-\frac{dP}{\cancel{dV}}\frac{\cancel{dV}}{dS}-P\frac{d^2V}{dVdS}\;.
\end{eqnarray}
Above equation, the Maxwell relation \eqref{MW1} is applied. Now we have
\begin{eqnarray}
    %\frac{\partial}{\partial S}\left(T\frac{dS}{dV}-P\right)&=&\frac{\partial}%{\partial V}\left(T-P\frac{dV}{dS}\right)\nn\\
    \cancel{\frac{dT}{dV}}+T\frac{d}{dV}\frac{\cancel{dS}}{\cancel{dS}}&=&-\cancel{\frac{dP}{dS}}-P\frac{d}{dS}\frac{\cancel{dV}}{\cancel{dV}}\;.
\end{eqnarray}
Again, above equation, the Maxwell relation \eqref{MW1} is applied. Then we see that the closure relation holds.
\end{proof}
\phantom{d}
\\
Naively, performing the variation with respect to dependent variables and imposing a condition $\delta U=0$, see appendix \ref{AppendixA}, we have the Euler-Lagrange equations as follows
\begin{eqnarray}
    \frac{\partial \mathcal{L}_V}{\partial S}-\frac{d}{dV}\left(\frac{\partial \mathcal{L}_V}{\partial\left(\frac{\partial S}{\partial V}\right)}\right)&=&0\;,\\
    \frac{\partial \mathcal{L}_S}{\partial V}-\frac{d}{dS}\left(\frac{\partial \mathcal{L}_S}{\partial\left(\frac{\partial V}{\partial S}\right)}\right)&=&0\;.
\end{eqnarray}
It is not difficult to see that if one substitutes the Lagrangian $\mathcal{L}_V$ and $\mathcal{L}_S$, the Maxwell relation \eqref{MW1} is trivially obtained. We note here that the explicit form of the Lagrangian like those in the context of the classical mechanics, namely $L\left( \frac{dq}{dt},q,t\right)=K\left(\frac{dq}{dt}\right)-V(q)$, may not relevant since one requires to derive first order differential equations, Maxwell relations, instead the second order differential equations.
\\
\\
We would like to point out that, upon working with the internal energy $U[S,V]$, there exists a $2d$ surface of the equilibrium states
\begin{equation}
    \Sigma=\Big\{(S,V,T,P): T=\frac{\partial \mathcal{L}_V}{\partial \left(\frac{\partial S}{\partial V}\right)}, \;-P=\frac{\partial \mathcal{L}_S}{\partial \left(\frac{\partial V}{\partial S}\right)};\; \frac{dT}{dV}=-\frac{dP}{dV}\Big\} \subset \mathbb{R}^4\;,\nn
\end{equation}
or equivalently
\begin{equation}
    \Sigma=\Big\{(S,V,T,P): T=\frac{\partial U}{\partial S},\; -P=\frac{\partial U}{\partial V};\; \frac{dT}{dV}=-\frac{dP}{dV}\Big\} \subset \mathbb{R}^4\;.\nn
\end{equation}
The path-independent feature of the thermodynamic evolution on this particular surface is captured through the Lagrangian 1-form closure relation.
\\
\\
We would also like to leave a final remark that, to obtain the Lagrangian closure relation, one can consider the differential form perspective. If one starts from the action remaining the same under the local deformation of the curve $\delta U=0$, we automatically have
\begin{equation}
    \oint_\Gamma \mathcal{L}_VdV+\mathcal{L}_SdS=\iint_A\left(\frac{\partial \mathcal{L}_V}{\partial S}-\frac{\partial \mathcal{L}_S}{\partial V}\right)dS\wedge dV+\left(\frac{\partial \mathcal{L}_S}{\partial V}-\frac{\partial \mathcal{L}_V}{\partial S}\right)dV\wedge dS=0\;
\end{equation}
with the assist of the Stokes theorem. Therefore, on the left hand side, two copies of the Lagarngian closure relation are presented. Moreover, on the right hand side, $(\mathcal{L}_V\;\;\mathcal{L}_S)^T$ can be treated as components of the vector field $dU=\mathcal L_VdV+\mathcal L_SdS$ given in figure \ref{F1}(b).

\section{Concluding summary}\label{SS}
We successfully formulate the action description to capture the reversible thermodynamic processes of the state function, namely the internal energy $U$. It turns out that a standard Lagrangian description is not appropriate to capture the exactness or the path-independent property, but one needs a Lagrangian 1-form description through a closure relation, or an integrability condition, as a direct result of the variational principle with respect independent variables. What we can draw a statement from this action principle for the equilibrium thermodynamics is that the change of the internal energy will follow the paths that $\delta U=0$, or $U$ is critical, from one equilibrium state to another equilibrium state. Of course, the idea can be directly employed to other thermodynamic potentials such as Helmholtz free energy, enthalpy and Gibbs free energy. We believe that this preliminary work would provide a first brick to lay out an alternative formulation of thermodynamics in the context of the classical mechanics.
\appendix
	\section{Derivation of the Euler-Lagrange equations} \label{AppendixA}
We shall proceed the steps given in\cite{Yookong1,Yookong2}. We first write the action functional
\begin{equation}
    U[S,V]=\int_{\mathscr{E}_\Gamma} \left(\mathcal{L}_V\left(\frac{dS}{dV},S;V\right)dV+\mathcal{L}_S\left(\frac{dV}{dS},V;S\right)dS\right)\;.
\end{equation}
Then we consider the variations: $S\rightarrow S+\delta S$ and $V\rightarrow V+\delta V$, resulting in
\begin{eqnarray}
    && U[S+\delta S,V+\delta V]-U[S,V]\nonumber\\
    &&=\int_{\mathscr{E}'_\Gamma}\left(\mathcal{L}_V\left(\frac{d}{dV}(S+\delta S),S+\delta S;V\right)dV+\mathcal{L}_S\left(\frac{d}{dS}(V+\delta V),V+\delta V;S\right)dS\right)\nonumber\\
    &&-\int_{\mathscr{E}_\Gamma} \left(\mathcal{L}_V\left(\frac{dS}{dV},S;V\right)dV+\mathcal{L}_S\left(\frac{dV}{dS},V;S\right)dS\right)\nonumber\\
    &&\approx \int_{\delta\mathscr{E}_\Gamma}\left(\frac{\partial \mathcal{L}_V}{\partial S}\delta S+\frac{\partial \mathcal{L}_V}{\partial \frac{dS}{dV}}\frac{d}{dV}\delta S \right)dV
    +\left(\frac{\partial \mathcal{L}_S}{\partial V}\delta V+\frac{\partial \mathcal{L}_S}{\partial \frac{dV}{dS}}\frac{d}{dS}\delta V \right)dS\;.
\end{eqnarray}
Integrating by parts and imposing the boundary conditions: $\delta S=\delta V=0$ at endpoints, one obtains
\begin{eqnarray}
    \delta U=U[S+\delta S,V+\delta V]-U[S,V]=\int_{\delta\mathscr{E}_\Gamma}\delta S\left(\frac{\partial \mathcal{L}_V}{\partial S}+\frac{d}{dV}\frac{\partial \mathcal{L}_V}{\partial \frac{dS}{dV}}\right)dV+\delta V\left(\frac{\partial \mathcal{L}_S}{\partial V}+\frac{d}{dS}\frac{\partial \mathcal{L}_S}{\partial \frac{dV}{dS}}\right)dS\;.
\end{eqnarray}
Therefore, the critical condition $\delta U=0$ leads to
\begin{eqnarray}
    \frac{\partial \mathcal{L}_V}{\partial S}+\frac{d}{dV}\left(\frac{\partial \mathcal{L}_V}{\partial \left(\frac{dS}{dV}\right)}\right)&=&0\;,\\
    \frac{\partial \mathcal{L}_S}{\partial V}+\frac{d}{dS}\left(\frac{\partial \mathcal{L}_S}{\partial \left(\frac{dV}{dS}\right)}\right)&=&0\;,
\end{eqnarray}
which are indeed the Euler-Lagrange equations.
%%%%%%%%%%%%%%%%%%%%%%%%%%%%%%%%%%%%%%%%%%%
\begin{acknowledgements}

S. Yoo-Kong would like to express a depth of gratitude to John Baez. Without his contribution to his blog, Azimuth, in the topic ``classical mechanics versus thermodynamics" part 1, 2 and 3, the idea on formulating the action principle in the present work would not be ignited.
\end{acknowledgements}
%%%%%%%%%%%%%%%%%%%%%%%%%%%%%%%%%%%%
%%%%%%%%%%%%%%%%%%%%%%%%%%%%%%%%%%%%%%%
%%%%%%%%%%%%%%%%%%%%%%%%%%%%%%%%%%%%

\end{document}